\documentclass[12pt,a4paper]{article}
\usepackage[dvips]{epsfig}
\newcommand{\ee}{$e^+e^-\: $}
\newcommand{\gaga}{$\gamma \gamma\:\;$}

\newcommand{\ttb}   {{\mathrm t\bar{\mathrm t}}}
\newcommand{\bb}    {{\mathrm b\bar{\mathrm b}}}
\newcommand{\cc}    {{\rm c\bar{\rm c}}}
\newcommand{\micron}{\mu\mathrm{m}}
\newcommand{\noi}{\noindent}

\begin{document}


\noi DESY 02 - 061

\vspace*{1.5mm}
\noi May 2002

\vspace{1cm}
\begin{center}
\LARGE{\bf SIMDET - Version 4\\
A Parametric Monte Carlo for a TESLA Detector}
\end{center}

\vspace{1.0cm}
\large
\begin{center}
M. Pohl\\Universit\'e de Gen\`eve, Switzerland, and\\
Katholieke Universiteit Nijmegen, The Netherlands\\
\vspace{3mm}
H.J. Schreiber\\ DESY Zeuthen, Germany \\
\end{center}

\vspace{2.0cm}

\normalsize
%
%
\begin{center}
\section*{Abstract}
\end{center}

A new release of the parametric detector Monte Carlo program
\verb+SIMDET+ (version 4.01) is now available.
We describe the principles of operation and the usage of this program
to simulate the response
of a detector for the TESLA linear collider. The detector components are
implemented according to the TESLA Technical Design Report.
All detector component responses are treated in a realistic way using a
parametrisation of results from the {\em ab initio} Monte Carlo
program \verb+BRAHMS+. Pattern recognition is emulated using a
complete cross reference between generated particles and detector
response. Also, for charged particles, the covariance matrix
and $dE/dx$ information are made available.
An idealised energy flow algorithm defines the output of
the program, consisting of particles generically classified as electrons,
photons, muons, charged and neutral hadrons 
as well as unresolved clusters. 
The program parameters adjustable by the user are described in detail.
User hooks inside the program and 
the output data structure are documented. 

\newpage

%
%
\section*{Introduction}

Recently the Technical Design Report (TDR) for the superconducting
linear collider TESLA~\cite{tdr} has been completed. A possible
detector as laid out in the TDR will have to deal
with a large dynamic range in energy, complexity of final states
and signal-to-background ratio. In particular, the
detector should be able
to cover the following important physics goals:

\begin{itemize}
\item very good momentum resolution 
      ($\delta 1/p_{T} \sim 4\cdot 10^{-5}/\mbox{GeV}$
      in the central region) to measure e.g. the Z recoil mass
      in the Higgsstrahlung
      process \ee $\rightarrow {\rm ZH}/{\rm Z}\rightarrow \ell^+ \ell^-$
      with optimal precision;
\item high resolution of hadronic jet energies
      ($\Delta E/E \approx 30\%/\sqrt{E}$)
      to reconstruct multi-jet events;
\item excellent b- and c-tagging capabilities to identify
      multi-b final states like ZHH or $\ttb$H and to separate
      ${\rm H} \rightarrow \bb, \, {\rm H} \rightarrow \cc$
      and ${\rm H} \rightarrow gg$;
\item good hermiticy to reduce background in missing energy channels; and
\item measurement capabilities in the forward direction.
\end{itemize}

The detector version implemented in \verb+SIMDET+ is based on
the description of the detector for the TESLA linear collider
as presented in Part IV of the TDR~\cite{tdr}. It has been
optimised for the physics requirements mentioned above. Its
basic components are:

\begin{itemize}
\item a multi-layer micro-vertex detector;
\item a tracker system with the main tracker as a large
      Time Projection Chamber (TPC), a
      silicon tracking subsystem between the
      vertex detector and the TPC
      and a set of forward chambers behind the TPC endplates;
\item a tracking electromagnetic calorimeter with very fine three dimensional
      granularity;
\item a hadronic calorimeter; 
\item and an instrumented mask with a small-angle luminosity detector.
\end{itemize}

\begin{figure}[htbp]
\begin{center}
\includegraphics[width=0.75\textwidth]{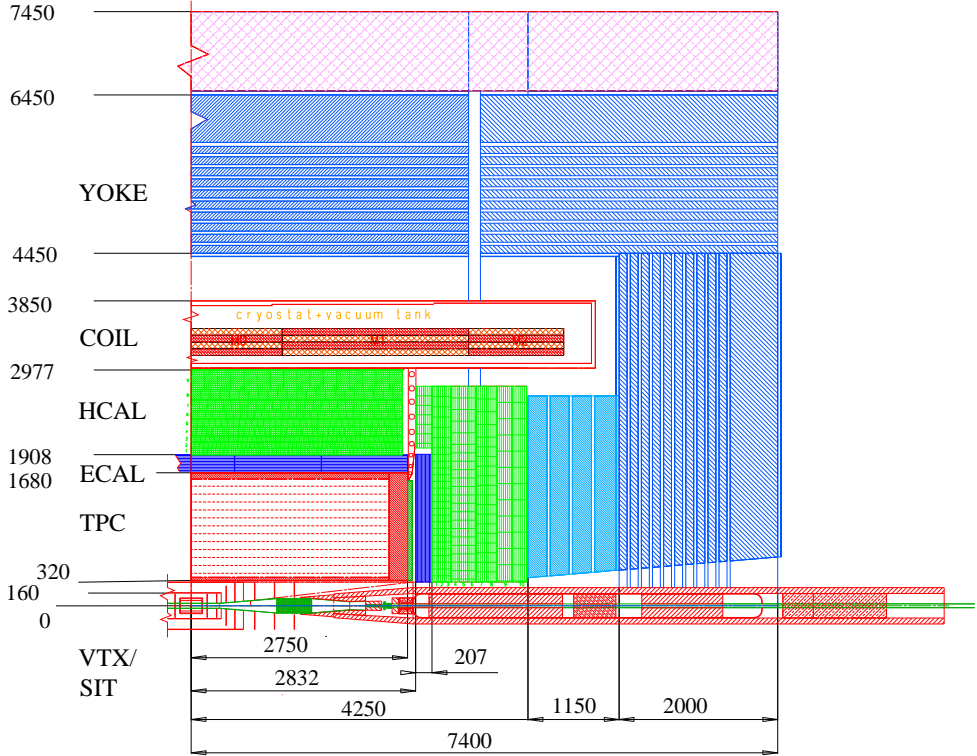}
\caption{View of one quadrant of the TESLA detector. 
  Dimensions ~are ~in ~mm.}
\label{tesla_detector}
\end{center}
\end{figure}

Fig.~\ref{tesla_detector} shows the principal layout
of the TESLA detector
and its dimensions. The tracking system and the calorimeters are
situated inside a 4 T superconducting solenoid.
\vspace{3mm}

At TESLA a bunch train of $\sim$ 2800 bunches with more than
300 ns spacing makes bunch identification easy and no special
detectors are needed for this purpose. Background expected is
mostly due to \ee pairs created in beam-beam interactions
produced at the interaction point (IP). These pairs
are concentrated at low polar angle and low transverse momentum.
In order to absorb background related to low-angle pairs a mask
in the forward direction is required which is also used for
calorimetry. The pairs at large angles can be kept at low radius
by a strong magnetic field. It limits the radius of the innermost
vertex detector layer to 1.5 cm which ensures an acceptable
background level.

\vspace{3mm}
The program \verb+SIMDET+ simulates the response of this detector
for linear collider physics in a fast and user-friendly manner.
The main objectives of the program are threefold:

\begin{itemize}
\item it provides physics studies groups with a tool for
      evaluation of the physics potential of a linear collider;
\item it treats the detector responds in a realistic manner using
      a parametrisation of results from the {\em ab initio} Monte Carlo
      program \verb+BRAHMS+;
\item it is fast and provides a simple output structure, so that
      physics studies can be done in a straight forward and
      decentralised way.
\end{itemize}

Only the {\em ab initio} GEANT3~\cite{geant} application program
\verb+BRAHMS+~\cite{brahms} provides detailed and realistic
information for charged particle momenta and directions,
impact parameters and calorimeter energies and directions.
Simple parametrisations of their resolution functions
are used as input for \verb+SIMDET+.
Gaussian smearing procedures then generate
the detector response. Starting from version 3 of the program,
a more accurate description of the calorimeter response  has been derived
from \verb+BRAHMS+ and a major upgrade
of the program was the implementation of an
energy flow algorithm to link tracker and calorimeter information. For
this purpose, a simple and idealistic cluster algorithm is applied to
the energy deposits in the calorimeters. Clusters and tracks are then
linked to form energy flow objects. The ultimate pattern recognition 
is emulated using a complete cross reference between the calorimeter
deposits and the particles originally generated. Energy flow objects are
then classified as electrons, photons, muons, charged and neutral
hadrons and clusters of unresolved particles
using the Monte Carlo truth information.
Also, if enabled, the covariance matrix for charged particles
and $dE/dx$ information are made available to the user. A new design of
the instrumented mask and a low-angle luminosity calorimeter have been
included into the default option of the detector,
so that the program is significantly enhanced as far as flavour
and particle identification capabilities and large angle
coverage are concerned.
The identification of a technology for the muon system
has yet to be explored in a comprehensive way. The program
\verb+SIMDET+ has been prepared for muon response simulation.
At present, however, muons are assumed to be measured by the
tracking device and identified applying a general misidentification
probability.

\vspace{3mm}
The objects produced by the
energy flow algorithm provide a consistent data structure at the
output of the program. This data structure, as well as defined user
hooks inside the program itself, can be used to perform physics
studies either within \verb+SIMDET+ or by means of an analysis program
after writing the energy flow objects to a file.

\vspace{3mm}
Furthermore, beamstrahlung based on parametrisations of Ref.\cite{Ohl}
can be included, \gaga hadronic background events can be overlayed
to physical events~\cite{Schulte}
and an event display for reconstructed objects has been
prepared~\cite{Vogt}. Recently, a \verb+CLIC+ version of the program
was initiated~\cite{Battaglia}
to study the physics potential of an \ee ~linear collider
in the center-of-mass energy region of 3 TeV and beyond.

\section*{Detector response simulation}

The anticipated particle physics program at TESLA, presented in detail
in Part III of the TDR~\cite{tdr}, represents a very demanding task
for the detector. With such a detector the sensitivity for
discovery and very high precision measurements
over energies from the Z peak up to $\sim$ 1 TeV
should be as large as possible, withstanding at the same time the
background expected. In this section we describe 
the response of the individual TESLA detector components
as implemented in the present version of \verb+SIMDET+.

\vspace{3mm}
Ongoing and planned R\&D activities for possible improvements or
redesigns of detector components are not considered in this note
and have to be accounted for in forthcoming \verb+SIMDET+ versions.

\subsection*{The tracker system}

\begin{figure}[htbp]
\begin{center}
\includegraphics[width=0.75\textwidth]{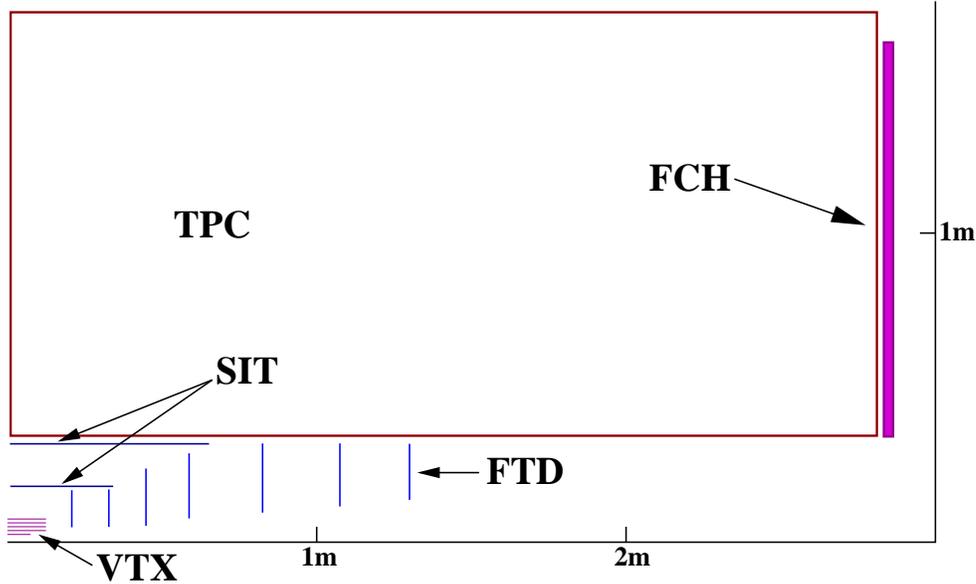}
\caption{The tracking system of the TESLA detector.}
\label{tesla_tracker}
\end{center}
\end{figure}

The concept of the tracking system
is shown in Fig.~\ref{tesla_tracker}.
Excellent tracking of charged particles is achieved by
a large Time Projection Chamber (TPC) as a main tracker,
supplemented by a silicon tracking detector between
the vertex detector (VTX) and the TPC, by discs in the
forward region and by a forward chamber located behind
the TPC endplate (FCH). The TPC design is similar to existing
ones, as implemented in the ALEPH or STAR detectors. So as to not to compromise
the momentum resolution and the calorimetric energy measurement,
field cage and endplates will be designed as thin as possible.
A TPC error of $\delta (1/p_{T}) \le 2\cdot 10^{-4}/\mbox{GeV}$
for large polar angle tracks and
$\sigma(dE/dx) \le 5 \%$ are anticipated. The intermediate tracker
consists of two layers of silicon strip detectors (SIT)
in the barrel region
down to $\theta = 25^\circ$  and three silicon pixel
and four silicon strip layers (FTD) on either side in the forward
region of the detector.
Single point resolutions of 10 $\micron$ respectively
50 $\micron$ are assumed in the simulations for these detectors.
The FCH, SIT and FTD improve significantly the
momentum resolution for all polar angles and, in particular, the angle
measurement capability in the very forward region
below $\theta = 12^\circ$. 

\vspace{3mm}
For the microvertex detector several technologies were proposed:
CCDs, active pixel sensors (APS) and CMOS sensors. With the CCDs a point
resolution of 3.5 $\micron$ and a layer thickness of 0.12 $\%$
of a radiation length can be reached. At present, the
CCD detector version has been implemented in detail in the
simulation program \verb+BRAHMS+. Consequently, the same information is
represented in \verb+SIMDET+.

\begin{figure*}[h!t]
\begin{center}
\mbox{\epsfxsize=17cm\epsffile{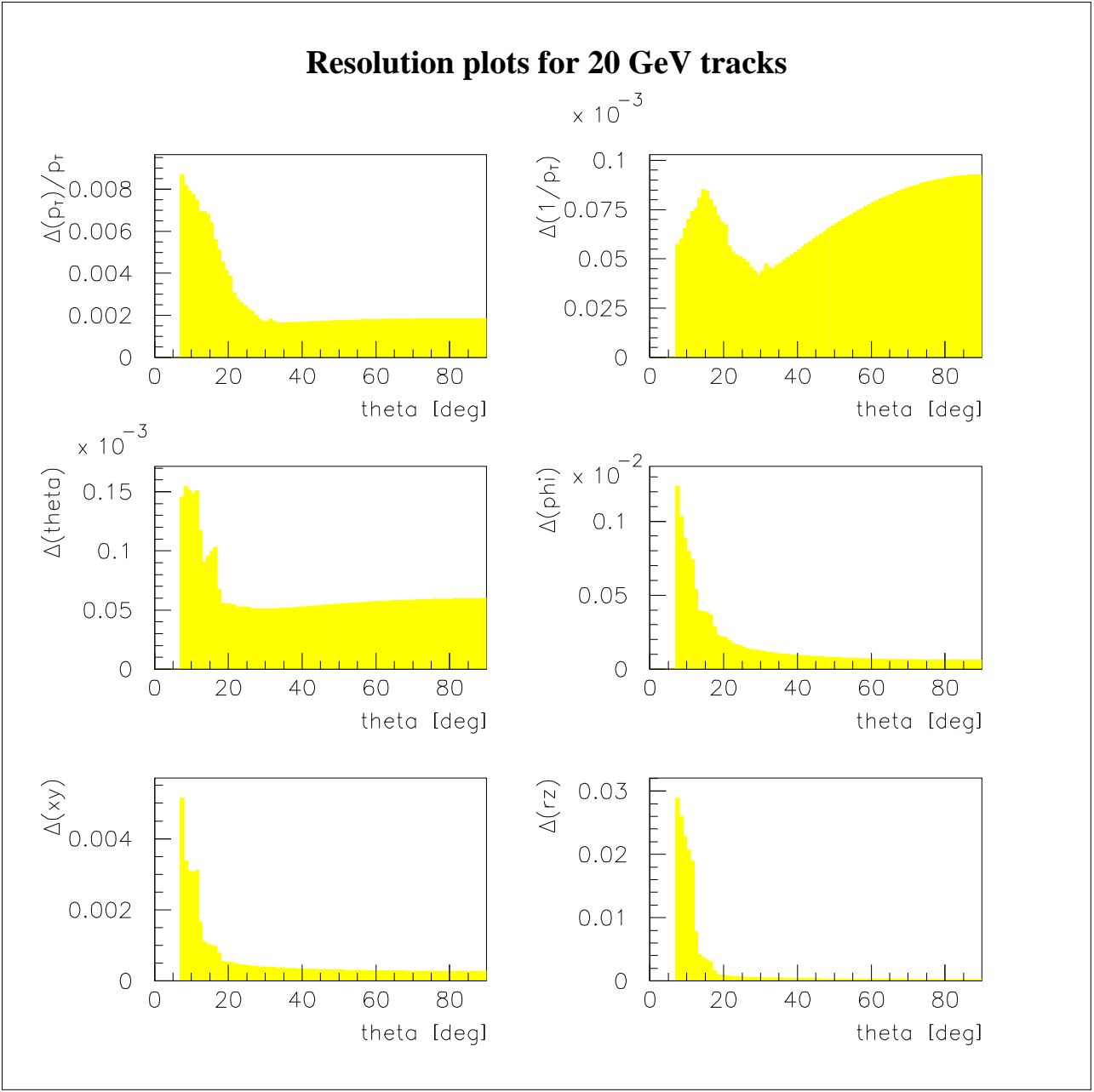}}
\end{center}
\vspace{-0.5cm}
\caption[ ]{\sl Resolution plots for 20 GeV momentum tracks as
 simulated with BRAHMS.}
\vspace{8mm}
\label{tesla_track_resol}
\end{figure*}

\vspace{3mm}
Using the overall tracking system, resolution functions for the inverse of the
transverse momentum, polar and azimuthal angles as well as impact parameters
as functions of momentum and polar angle $\theta$
are input to Gaussian smearing procedures.
Interpolating between data points and extrapolation to not yet 
simulated regions, reasonable descriptions in the
whole $(p,\theta$)-plane have been obtained. Attention has been
paid to ensure that the parametrised resolutions are in agreement
within a few percent with the \verb+BRAHMS+ predictions. The inverse transverse
momentum, the directions $\theta$ and $\phi$ and the impact parameter
resolution parameters are smeared independently. Correlations between
momenta and $\phi$ directions of tracks are neglected.
An example of the \verb+BRAHMS+ simulated resolution functions
for 20 GeV momentum tracks are shown in Fig.~\ref{tesla_track_resol} as a 
function of polar angle.

\vspace{3mm}
The measured values thus obtained for 1/$p_{T}, \theta$ and
$\phi$ can be transformed to $p_{x}, p_{y}$ and $p_{z}$ by means
of the program package described in Ref.~\cite{blobel}, including 
proper propagation of the measurement errors.

\vspace{3mm}
\verb+SIMDET+ offers to set a minimum transverse momentum for tracks,
so that below that value charged particles are removed from the
event. Tracking efficiency and charge misinterpretation
probability can also be specified. The track reconstruction efficiency
is momentum dependent: it equals the (default) data card value
for tracks with momentum above 2 GeV and becomes
continuously smaller with decreasing track momentum.
Charge misinterpretation according to the (default) data card value
is only applied to large $p_{T}$ tracks; this value is
increased by a factor of three for badly measured particles.
No attempt is currently made to take
into account overlaps of closely spaced tracks.


\vspace{3mm}
The main element
entering in the lifetime tagging is the impact parameter.
The impact parameter in the $(r-\phi)$ plane is defined as the minimal
distance between the primary vertex (PV) and the track trajectory
projected onto the plane perpendicular to the beam direction.
The point of closest approach ($P_{C}$) of the track trajectory
to the primary vertex in the $(r-\phi)$ plane is also used to define
the $(r-z)$ projection of the impact parameter.
Then the $(r-z)$ projection of the track
impact parameter is the difference between the z-coordinates
of the primary vertex and the point $P_{C}$.

\vspace{3mm}
Both impact parameters have a sign defined as follows.
The point of closest approach of the track to the simulated
flight direction is computed and the negative (positive) sign
is set if this point is upstream (downstream) with respect to
the primary vertex position. With this definition, tracks from
decays of long-lived particles have predominantly positive signs
while tracks coming directly from the PV are equally likely
to be positive or negative due to the limited precision
of track reconstruction.

\vspace{3mm}
For the CCD version of the vertex detector,
all necessary input for sophisticated flavour identification exists
in the form of a parametrised covariance matrix.


\vspace{3mm}
For the APS version of the vertex detector, GEANT3 simulations provide
the impact parameter errors as follows:

\begin{eqnarray}
  \sigma_{r\phi} & = & A \oplus \frac{B}{p\sin^{\frac{3}{2}}\theta}\\
  \sigma_{rz} & = & C \oplus \frac{D}{p\sin^{\frac{5}{2}}\theta}
\end{eqnarray}
for a track with momentum p and polar angle $\theta$
in the $(r-\phi)$ and the $(r-z)$ projections. The values of the
parameters $A$, $B$, $C$, and $D$ were obtained from \verb+BRAHMS+
simulations and transmitted to \verb+SIMDET+.

\subsection*{Calorimetric response and energy flow}

The physics programme at a linear collider calls for calorimetry
with unprecedented performance which can be translated into the
following requirements:

\begin{itemize}
\item hermeticity down to very small angles,
\item excellent energy resolution for jets respectively partons,
\item excellent angular resolution,
\item capability to reconstruct non-pointing photons as a
      stand-alone device,
\item good time resolution, to avoid event pile-up.
\end{itemize}

Dense and hermetic sampling calorimeters with very high
granularity realise at best these demands. In the TDR, two options
are presented both for the electromagnetic (ECAL) and the
hadronic (HCAL) calorimeter components.
For the ECAL, the two options are
a very high granularity 3D calorimeter based on tungsten and
silicon diode pads and a shashlik (scintillator/lead)
calorimeter as presented
in the CDR~\cite{cdr}. Both versions have undergone
a successful R\&D program,
detailed layout studies and careful implementations into the
simulation program \verb+BRAHMS+.

\vspace{3mm}
For the HCAL, the two suggestios are an Fe/scintillating tile
calorimeter with fine transverse and longitudinal segmentations
and a fully digital calorimeter, which represents a novel approach
to hadronic calorimetry. It consists of fine pixels
(1$\times1 cm^{2}$) for an iron/gas sandwhich that is read-out
digitally (yes/no) for each cell.

\vspace{3mm}
In Version 4 of \verb+SIMDET+, the tungsten/silicon option
for the electromagnetic component and the Fe/scintillating tile option
for the HCAL are implemented.

\begin{figure*}[htbp]
\begin{center}
\mbox{\epsfxsize=0.49\textwidth\epsffile{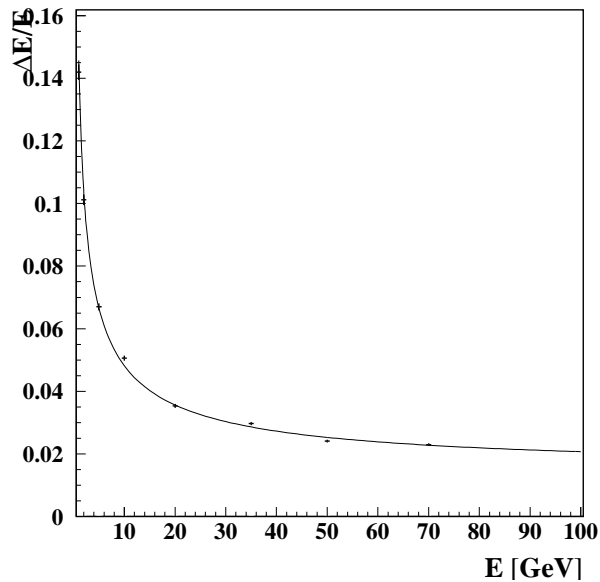}}
\end{center}
\vspace{-0.5cm}
\caption[ ]{\sl Relative energy resolution as a function of energy for single 
high energy electrons as simulated with BRAHMS.}
\label{fig:egamres}
\end{figure*}

\begin{figure*}[htbp]
\begin{center}
\mbox{\epsfxsize=0.49\textwidth\epsffile{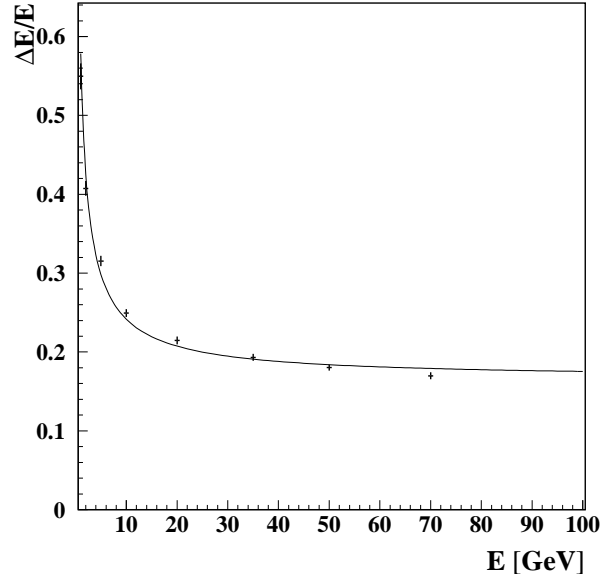}}
\end{center}
\vspace{-0.5cm}
\caption[ ]{\sl Relative energy resolution as a function of energy for single 
pions as simulated with BRAHMS.}
\label{fig:hadres}
\end{figure*}

\vspace{3mm}
The simulation of the calorimetric response to hadrons and leptons has been 
updated to correspond more closely to the expected performance of the TESLA 
calorimeters as described in the Technical Design Report. The excellent 
energy resolutions for single electrons and photons at high energies are
shown in Fig.~\ref{fig:egamres}, the energy resolution for hadrons in
Fig.~\ref{fig:hadres}. Both resolution curves have have been determined
using {\em ab initio} simulations of the baseline electromagnetic and hadronic 
calorimeters as implemented in \verb+BRAHMS+. They agree well with the 
resolutions quoted in the TESLA Technical Design Report\cite{tdr}.
These resolution curves are implemented in the current version of 
\verb+SIMDET+. 

On the other hand, no attempt has been made to simulate
the benefits expected from the extremely fine three-dimensional granularity
of the silicon tungsten calorimeter. Consequently, also the subsequent cluster
algorithm sticks to the simple two-dimensional tower geometry used in the 
previous version of the program. Users who require a more accurate
simulation of the spatial energy deposit or more sophisticated pattern
recognition in the calorimeters should directly use \verb+BRAHMS+ for
their study.

\begin{figure*}[htbp]
\begin{center}
\mbox{\epsfxsize=0.49\textwidth\epsffile{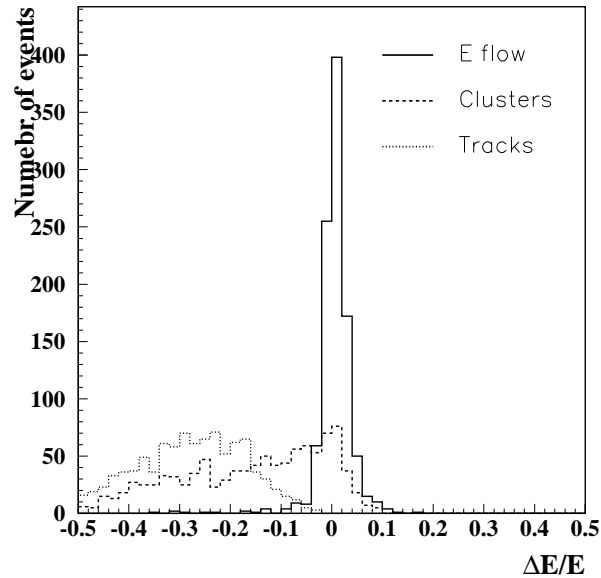}}
\end{center}
\vspace{-0.5cm}
\caption[ ]{\sl Energy resolution, using different estimators, for the process
$\mathrm{e^+ e^-} \rightarrow \mathrm{H Z}$ at 500 GeV center of mass energy, 
where the Z boson decays hadronically, while the Higgs boson decays into two W
bosons which in turn decay leptonically. The histograms represent the
relative difference between true and measured energy using
tracks alone (dotted line), calorimetric clusters alone (dashed line)
and the energy flow (full line).}
\label{fig:eflow}
\end{figure*}

\vspace{3mm}
The energy flow output implemented in the present version of \verb+SIMDET+
which was left unchanged with respect to Version 3~\cite{version3},
thus does not
fully benefit from the superb pattern recognition capabilities of the 
tracking calorimeter. Nevertheless, as demonstrated in Fig.~\ref{fig:eflow},
usage of the energy flow concept leads to a vast improvement in the 
reconstruction of the visible and missing energy.

\subsection*{Instrumented mask and forward luminosity calorimeter}

The standard calorimeters ECAL and HCAL are completed by
forward calorimeters, which cover polar angles down to 4.6 mrad.
These calorimeters, despite their small size, have a large impact
on the overall detector performance, since they enhance missing
energy resolution, provide electron/photon identification
and measure single bunch luminosity.

\vspace{3mm}
The Low Angle Tagger (LAT) is designed at the tips of the tungsten
mask with a $\theta$ coverage down to $\sim$30 mrad and serves as an
additional shield to protect the tracking detectors from
backscattered particles. Its design foresees a tungsten sampling
calorimeter. It is supported by an inactive
tungsten structure. \verb+BRAHMS+ simulations assume
radial and azimuthal segmentations for this device.
The background expected
from \ee beamstrahlung pairs is comparatively small, so that
the LAT can probably be used for (electron, photon) and muon
identification and measurement.

\begin{figure*}[h!t]
\begin{center}
\mbox{\epsfxsize=14cm\epsffile{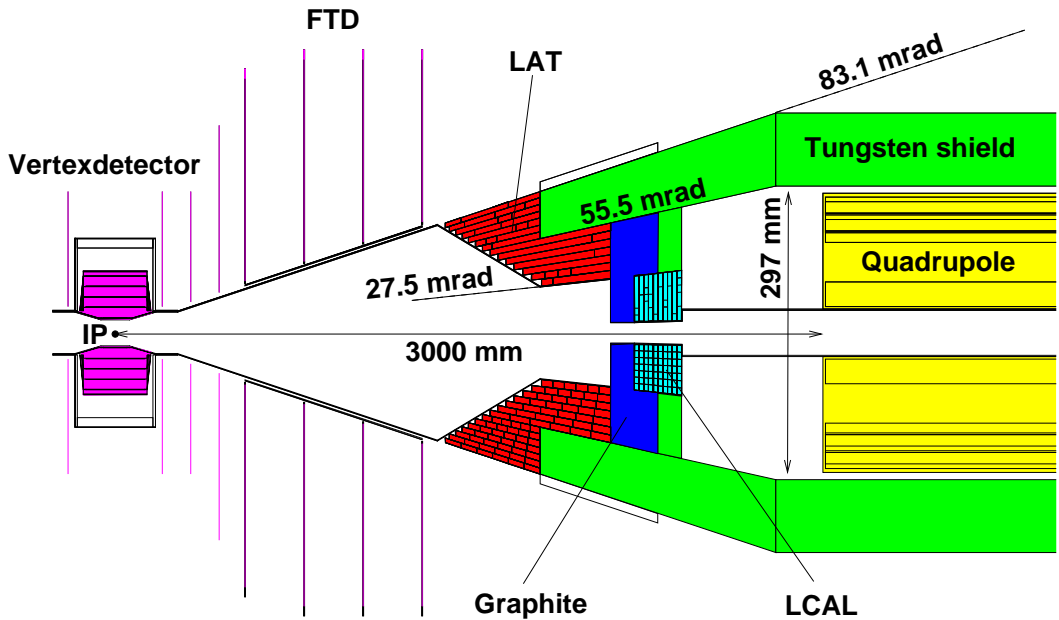}}
\end{center}
\caption[ ]{\sl Design of the forward angle calorimeters,
 LAT and LCAL.}
\vspace{4mm}
\label{tesla_LAT_LCAL}
\end{figure*}

\vspace{3mm}
The Low Angle Calorimeter (LCAL) serves both as a fast luminosity
monitor and as a low angle calorimeter. The current design
foresees a sampling calorimeter with segmentation in the $z$ direction
and azimuthal subdivision. It has to withstand a high level of
electromagnetic radiation, yet render accurate energy
and angular resolutions. In Fig.~\ref{tesla_LAT_LCAL} the placement
of the two calorimeters in the mask structure is shown.


\vspace{3mm}
The expected performance of the LAT has been studied using
\verb+BRAHMS+ responses for high energy electrons and muons.
Energy and $\theta$ resolution functions for these particles
were parametrised and the parameters obtained are
implemented in \verb+SIMDET+.
Also, detection efficiencies as functions of energy
and polar angle are taken into account.
Since no charge information
exists, electrons and positrons are interpreted as objects with
unknown charge and treated as photons.
Closely spaced $e^+$, $e^-$ and photons
are merged to a new energy flow object, while
muons with sufficient energy are considered to be
isolated and are given a random charge.
\verb+SIMDET+ allows to overwrite the default energy and angular
resolution parameters as well as the minimum energy needed for particle
detection. 

\vspace{3mm}
For the LCAL we expect from existing \verb+BRAHMS+ simulations that only
$e^\pm$ and photons will be measurable with energy
and polar angle dependent probabilities. 
The resolutions expected for energetic particles in light
of intense backgrounds
are taken into account in \verb+SIMDET+. All measured LCAL objects
are interpreted as photons, since charge information is missing.
 Accounting for some
minimum energy deposits it seems possible that the detector
allows electron measurements with more than 50 GeV down to polar angles
of 5 mrad.

Since the LCAL is at present only in a stage of preliminary design,
the current \verb+SIMDET+ response simulation is simple
and accounts only for relative restricted information
from the full Monte Carlo program \verb+BRAHMS+.

%
\section*{Particle identification with $dE/dx$}

The $dE/dx$ simulation is based on the Bethe-Bloch equation which is a
universal function of $\beta\gamma$ for all particle species. The
expected $dE/dx$ for a track is generated according to its mass
and momentum using the Bethe-Bloch equation with parameters derived from
the OPAL jet chamber. Then, the expected mean is smeared by an error
that depends on the number of isolated samples, on the sampling length
and on the error of the sampling length that results from the uncertainty
of the track polar angle $\theta$.
 Parameters of the $dE/dx$ error description
are based on a study with a dedicated Monte Carlo simulation ~\cite{hauschild}
where individual electrons
and clusters have been generated and tracked to the TPC endplate
including diffusion, gas gain fluctuations and crosstalk.

The $dE/dx$ code provides for the electron, muon, pion, kaon and proton
hypothesis five signed weights of range -1 to +1. The
five weights are the signed $\chi^{2}$ probabilities of a track being
consistent with a certain particle species. The sign of the weight
corresponds to the sign of the difference $dE/dx_{\mbox{meas}} -
dE/dx_{\mbox{exp}}$, where $dE/dx_{\mbox{exp}}$ is the $dE/dx$ value 
according to the Bethe-Bloch equation and $dE/dx_{\mbox{meas}}$ is the 
expected $dE/dx$ smeared with the appropriate error.

The $\chi^{2}$ probabilities also include the track momentum error
and are calculated by minimising the following 
two-dimensional variable:

\vspace{3mm}
 $\chi^{2} = \{(dE/dx_{meas} - dE/dx_{exp})/\sigma(dE/dx)\}^{2}
           + \{ ( p_{meas} - p_{exp} ) / \sigma(p)\}^{2}$ \\

\vspace{1mm}
The signed weights also allow the conversion into
equivalent normalised $dE/dx$ pull values.
For particle identification purposes either the "probability
frame" might be used where cuts on $dE/dx$ weights are placed or the
"normalised $dE/dx$ frame" could be used where cuts on some standard
deviations are placed. However, there is no general preference for one or
the other framework as both are fully equivalent.

\vspace{3mm}
Please note that the five $dE/dx$ weights in general do not
necessarily add up to one. For two different particle species,
at a certain momentum, the expected $dE/dx$ values from the Bethe-Bloch
equation are identical and indistinguishable. At this cross-over point,
the probabilities for both particle hypotheses
are exactly equal to one.

\vspace{3mm}
The expected particle separation power depends
on the amount of the isolated samples available per track.
Usually tracks are imbedded in dense track environments
(multi-hadronic jets), which reduce the number of samples significantly
due to the limited double track resolution of the TPC.
The program \verb+SIMDET+ allows to estimate the amount of overlap regions
for each track by means of two-particle separation distances
in the $(r-\phi)$ and $(r-z)$ projections. Since each track is
compared with all others, speed of computation can be governed
by a step size parameter and the possibility to cut-off the
comparison of tracks after a certain number of consecutive space points
taking into account the spread-out of charged particles
within the magnetic field.

\section*{Usage notes}

 \textbf{SIMDET}
 has a built in an interface to \textbf{PYTHIA}
 for internal event generation via subroutine \textbf{SIPYTH}.
 Alternatively, events generated by other programs can be used 
for detector simulation; subroutine \textbf{SIEVTI}
 provides an input interface to events written by 
subroutine \textbf{LCWRITE}.
If the structure of the generated events does not coincide
 with common \textbf{/PYJETS/} of \textbf{PYTHIA},
subroutine \textbf{SIPREP} should fill the arrays 
\textbf{K$_{-}$PY(I,...)},
\textbf{P$_{-}$PY(I,...)} and \textbf{V$_{-}$PY(I,...)}.

To allow for initial state radiation and beamstrahlung during event
generation is in the responsibility of the event generator.
Subroutine \textbf{SIPYTH} accounts for both radiations if wanted.

In all cases, \textbf{NEVENT}
 specifies the number of events to be processed. 

The detector parameters used are defined in subroutine \textbf{SIDETR}.
The naming convention is self explanatory. 

  After selection of all particles which might give a response
 in the detector (subroutine \textbf{SIPART}),
charged particle tracking, impact parameter estimations and the
covariance matrix are obtained from the subroutine \textbf{SITRAK},
subroutine \textbf{SICDAS} and subroutine \textbf{SICOVM}, respectively.

Detector resolution parameters for tracks are provided by the
ab initio Monte Carlo program \textbf{BRAHMS} taking into
account the complete tracking system.

\vspace{3mm}
Two options exist for the vertex detector (subroutine \textbf{SIDCAS}):

\begin{enumerate}
\item the APS version, based on 1.5 cm beam pipe radius
\item the CCD version, based on 1.5 cm beam pipe radius
\end{enumerate}
and only the CCD version allows for the covariance matrix at present.

The calorimetric response is also based on detector studies
using the {\em ab initio} Monte Carlo simulation package
\textbf{BRAHMS}.
The energy distributions of electrons and charged pions 
in adjacent cells were fitted and the parameters obtained
 are used for appropriate energy deposite simulations
in the electromagnetic and hadronic calorimeters 
(subroutine \textbf{SILEGO} and subroutine \textbf{SIDEPO}).
 Afterwards, a cluster search algorithm (with some 
idealised assumptions)
 provides clusters (subroutine \textbf{SICLUS\_I})
in the ECAL as well as the HCAL. Finally, an energy flow algorithm
 (subroutine \textbf{SIFLOW\_I})
joins tracker and calorimeter informations such that 
as many single particles as possible are resolved. 
At this stage, the best estimate for charged particles
is simply taken from the tracker,
except for high energy electrons and positrons.
Here, a proper weighting scheme of their tracker and calorimeter energies
is applied (subroutine \textbf{SIWGHT}).
Particle identification on the basis of shower shapes 
and matching between tracker and calorimeter information 
in space and momentum is emulated. Finally, best estimates 
of the energy flow objects are established.

In the very forward direction an instrumented mask,
the low-angle tagger (subroutine \textbf{SILOWT}),
and a low-angle luminosity calorimeter (subroutine \textbf{SILCAL}) are
included as default devices. Their preliminary responses 
from \textbf{BRAHMS} simulation are included in \textbf{SIMDET}.

Particles not entering the calorimeters, e.g. low-pt tracks,
are rescued in subroutine \textbf{SIRSCU}
and added to the list of energy flow objects in
subroutine \textbf{SISTOR}.

In addition, \textbf{SIMDET} allows for particle identification
from specific energy losses $dE/dx$ within the TPC
(steered by subroutine \textbf{SIDEDX}).

Background events of the type \gaga $\rightarrow$ hadrons (and
later also of other sources) can be overlayed to each physical
event using the package \textbf{hades}~\cite{Schulte}. The data card
\textbf{IBKG} enables this option provided the name of the
background file (data card \textbf{BKGF}) and the average number
of background events to be overlayed (data card \textbf{NBKG})
are given by the user. Also the study of only background events
is possible. Formatted background files
resulting from  the package \textbf{GUINEA$_{-}$PIG}
exist for the \textbf{TESLA} collider at cms energies
170, 360, 500 and 800 GeV as well as for two \textbf{CLIC} options:
the nominal \textbf{CLIC} parameter option with a luminosity
1.05 $\dot10^{35}$ cm$^{-2}$sec$^{-1}$ at 3 TeV and an optimised version with
lower beamstrahlung and a luminosity of 0.4 $\dot10^{35}$ cm$^{-2}$sec$^{-1}$.
It is the responsibility of the user to transform these files to
a local binary file using the program \textbf{hades$_{-}$import}.

The \textbf{CLIC} option of \textbf{SIMDET} can be enabled
by the data card \textbf{CLIC}. If it equals 1, the process
selected in subroutine \textbf{SIPYTH} is simulated at the
predetermined cms energy, while \textbf{CLIC 2} allows for
beamstrahlung effects if a luminosity spectrum file is provided on the
data card \textbf{CLCF}. This option reads the
electron and positron energies as obtained from \textbf{GUINEA$_{-}$PIG}
for a nominal beam energy of 1500 GeV. If a different energy
is desired, the simulated \ee energies are rescaled. In the case
of invoking a luminosity spectrum file, note that the file name
extension '.info' respectively '.ep' should not be given on the
data card \textbf{CLCF}. For more details we refer to~\cite{Battaglia}.

The history of the events can be monitored
either completely or in a restricted
form. If the data card \textbf{HISTORY} is enabled, the
complete event history is stored as described in the
\textbf{generated particle record}, otherwise input quantities of
only stable particles are kept, with a flag of being
accepted by the detector or not.

Booking of histograms is provided in subroutine \textbf{SIBOOK}.
A few standard histograms allow to monitor the detector response.
They can be optionally filled and switched off or on by
the data card \textbf{PLOT}.

The simulation finishes by providing all measured objects 
in the proposed common output structure in subroutine \textbf{SISTOR}.

If requested, the program allows to write (unformatted/formatted)
 all objects in either the standard structure or in 
a restricted structure (best estimates only) to an 
external file in subroutine \textbf{SIWDST}
 so that analysis can follow externally. It also offers to produce
 zipped output files (patchy use selection \textbf{GZIO} and
data card \textbf{GZIO}) in order to save disk space.
If this option is chosen, the extension \textbf{.gz} is 
added by the program to the file name.
The required package is provided with the \textbf{SIMDET} code.

 If no external file should be written, the array \textbf{VECP(I,K)}
 is optionally filled in subroutine \textbf{SIVECS}
 which allows to use directly the physics analysis package
 \textbf{VECSUB} (a product from SLAC/DELPHI). In this case, the
components 1 to 7 of the array \textbf{VECP} involve the
quantities as described in the \textbf{VECSUB-DELPHI} note ~\cite{vecsub},
 while the originally free
locations 8, 9 and 10 of this array contain now the
transverse and the longitudinal impact parameters, in units of sigma,
and a simple numbering of the energy flow objects accepted.

  Subroutine \textbf{SIFFRE}
 reads the set of data cards with the \textbf{FFREAD}
 package. Default settings in subroutine \textbf{SIINIT}
 and/or in subroutine \textbf{SIDETR}
 might be overwritten. 

 A user subroutine \textbf{SIUSER (IFLAG)} has been added.
It is called during inititialization (IFLAG=1), 
for each event before writing to output file (IFLAG=2)
 and during termination of the job (IFLAG=3). 

\vspace{3mm}
The package exists as a \textbf{CVS} repository
as well as \textbf{PATCHY/CMZ CAR}
files. If the \textbf{PATCHY} distribution is chosen,
the desired code generation is steered
by 'use' selections. For the \textbf{CVS} repository
the selections are given as arguments to the
configure script.

\textbf{SIMDET} selects the standard simulation code,
and with \textbf{NOSIMU}
only all stable particles without any detector response are treated.
This enables physics studies at the parton or generator level.

\vspace{3mm}
Further \textbf{PATCHY} 'use' options are:\\

\textbf{CIRCE} to allow for the beamstrahlung code,\\

\textbf{COVMTX} for the covariance matrix material,\\

\textbf{GZIO} to produce zipped output files,\\

\textbf{IBKG} to invoke additional code for background events and \\

\textbf{CLIC} for the CLIC linear collider option.

\vspace{3mm}
Please note that for HP platforms error handling routines are
included (forced by the 'use' selection \textbf{HPUX}).

\vspace{3mm}
Please link the \textbf{CERNLIB} libraries, the \textbf{PYTHIA 6.1}
library and, if selected, the enclosed \textbf{GZIO} library,
as transparently illustrated by the installation script which can
be used for different platforms (Linux, HP-UX, SunOS, OSF1).

\vspace{3mm}
A small utility package \textbf{SIANAL} has been added which can serve
as a template for further analyses. It contains routines to open and
read the \textbf{SIMDET} output file of the energy flow objects
written formatted or unformatted. For both types also
zipped files can be read. A zipped file is assumed if the file name has
the extension '.gz'. The file name is given in the command line to run
\textbf{SIANAL}, eg. '$./sianal.bin$ $simmdet\_v4.evt.gz$'.

\subsection*{Input/Output units:}
 
\begin{tabular}{lll}
 Logical unit number  & Default file name  & Contents  \\
 6  & simdet$v_{-}$4.res  & debugging information like the history \\
    &                & of a few events and histograms  \\
 7  & user-defined  & input events generated by an external program  \\
 8  & simdet$v_{-}$4.dat  & free format data card file  \\
 9  & user-defined  & input \gaga background events \\
    &               & generated by an external program  \\
 10 & user-defined  & CLIC luminosity spectrum file \\
    &               & generated by an external program  \\
 11  & user-defined  & simulated and reconstructed objects for further analysis  \\
    &                & (for zipped output the extension '.gz' is added\\
    &                &  automatically) \\
 12  & simdetv$_{-}$4.hist  & PAW histogram file \\

\end{tabular}

\vspace{3mm}
  Note that the names of input and reconstructed event files
as well as \gaga background event and the CLIC luminosity
spectrum files
 should be enclosed in single quotes, start with a \textbf{point}
 and end with a \textbf{blank}
 when specified. The length of all file names is restricted to 80 characters.

\section*{Program parameters and steering}


In the appendices to this description, the main program parameters,
user accessible COMMON blocks, data cards and the output structure
are described in detail. An up-date description together with advice
for their usage is also given in the \verb+SIMDET+ online documentation
at \\
$http://www.ifh.de/linear_{-}collider/users_{-}guide_{-}simdetv4.html$

\section*{Acknowledgements}

The authors would like to express their gratitude to
M. Battaglia, K. Buesser, C. Damerell, \linebreak
V. Djordjadze, P. Garcia-Abia,
M. Hauschild, K. Moenig, Th. Ohl, D. Schulte, N. Tesch and \linebreak
H. Vogt
for important contributions to the present version of \verb+SIMDET+
and valuable discussions. We also acknowledge comments and suggestions
from many people too numerous to be mentioned,
which improved the program in many respects.
We would also like to thank V. Blobel for providing his program package
for data analysis.

\newpage
\section*{Appendix A: detector parameters}

The parameters of the detector are specified in subroutine \verb+SIDETR+
and subroutine \verb+SIFFRE+. They are printed in subroutine \verb+SIPRNT+
for user control. Their meaning is described in the following:

\vspace{3mm}
\begin{tabular}{ll}
 FIELD   &  field of the solenoid [Tesla] \\
         & \\
 COSCCD  &  cos(theta) polar angle acceptance value of the CCD
                        vertex detector \\

 COSAPS  &   cos(theta) polar angle acceptance value of the APS
                        vertex detector \\
         & \\
 TPCIRAD &  inner radius of the TPC [m] \\
 TPCORAD &  outer radius of the TPC [m] \\
 TPCLEN  &  total lenght of the TPC [m] \\
 COSTPC  &  cos(theta) polar angle acceptance value of the TPC \\
 COSTRK  &  cos(theta) polar angle acceptance value of the
                        total tracking system \\
 PTMINP  &  minimum transverse momentum of a track accepted
                        for tracking [GeV] \\
 TRKEFF  &  track reconstruction efficiency \\
 QTRKMIS &  charge misinterpretation probability \\
         & \\
 COSECAL &  cos(theta) polar angle acceptance value of the
                        electromagnetic calorimeter \\
 EMECAL  &  minimum deposited energy required in the
                        electromagnetic calorimeter [GeV] \\
 AEMRES  &  stochastic parameter for the energy resolution
                        in the electromagnetic calorimeter \\
 BEMRES  &  constant parameter for the energy resolution
                        in the electromagnetic calorimeter \\
 EPIMIS  &  electron misinterpretation probability \\
         & \\
 COSHCAL &  cos(theta) polar angle acceptance value of the
                        hadronic calorimeter \\
 EMINHA  &  minimum deposited energy required in the
                        hadronic calorimeter [GeV] \\
 AHARES  &  stochastic parameter for the energy resolution
                        in the hadronic calorimeter \\
 BHARES  &  constant parameter for the energy resolution
                        in the hadronic calorimeter \\
         & \\
 COSCAL  &  cos(theta) polar angle acceptance value of
                        the total calorimeter \\
         & \\
 COSLLAT &  min cos(theta) polar angle acceptance value of the
                        low angle tagger \\
 COSULAT &  max cos(theta) polar angle acceptance value of the
                        low angle tagger \\
 EMINLAT &  minimum deposited energy required in the
                        low angle tagger [GeV] \\
 ALATRES &  stochastic parameter for the energy resolution
                        in the low angle tagger \\
 BLATRES &  constant parameter for the energy resolution
                        in the low angle tagger \\
 CLATRES &  cell size in theta of the low angle tagger [rad] \\
 DLATRES &  cell size in phi of the low angle tagger [rad] \\
         & \\
 COSLLAC &  min cos(theta) polar angle acceptance value of the
                        low angle calorimeter \\
 COSULAC &  max cos(theta) polar angle acceptance value of the
                        low angle calorimeter \\
         & \\
 EMUDEP  &  averaged deposited energy
                        of muons in the ECAL and HCAL [GeV] \\
 EMUISO  &  min energy of muons considered of being isolated [GeV] \\
 XMUMIS  &  muon misidentification probability \\
         & \\
\end{tabular}

\newpage
\section*{Appendix B: User accessible COMMON blocks}

The most important user accessible COMMON blocks are
shortly described in the following:

\vspace{3mm}
\begin{verbatim}
+KEEP, SIEVNT.
       COMMON /SIEVNT/ NEVENT, IEVENT, IEOTRI, IEORUN, IDCHAN, ECMS
      +,               BKGEVT, XSECTION, CMHEAD(10)
       INTEGER         ICMHEAD(10)
       EQUIVALENCE     (ICMHEAD(1), CMHEAD(1))
*
\end{verbatim}
\vspace{-4mm}
\begin{tabular}{ll}
NEVENT   &  Number of events to be processed \\
IEVENT   &  Current event sequence number \\
IEOTRI   &  Flag to abort current event of non zero \\
IEORUN   &  Flag to terminate run of non zero \\
IDCHAN   &  Reaction channel identificator \\
ECMS     &  c.m.s. energy [GeV] \\
BKGEVT   &  average number of background events to be overlayed to a physical event \\
XSECTION &  cross section [fb] \\
CMHEAD   &  Header of files (10 words, referenced in App. D and E) \\
\end{tabular}

\vspace{3mm}
\begin{verbatim}
+KEEP, SIUNIT.
       COMMON /SIUNIT/ LUNIN, LUNOUT, LUNDST, LUNHIS, LUNEVT, LUNDAT
      +,               LUNBKG
*
\end{verbatim}
\vspace{-4mm}
\begin{tabular}{ll}
LUNIN    &  Standard input unit \\
LUNOUT   &  Standard output unit \\
LUNDST   &  Unit to write reconstructed event information \\
LUNHIS   &  Unit for histograms \\
LUNEVT   &  Unit to read generated events \\
LUNDAT   &  Unit for free format data cards \\
LUNBKG   &  Unit to read \gaga $\rightarrow$ hadrons background events \\
\end{tabular}

\vspace{3mm}
\begin{verbatim}
+KEEP, SICONS.
       COMMON /SICONS/ PI, TWOPI, PIBY2, SMALL, RADDEG, DEGRAD
*
\end{verbatim}
\vspace{-4mm}
\begin{tabular}{ll}
PI       &  Number PI (ACOS(-1.)) \\
TWOPI    &  2*PI \\
PIBY2    &  PI/2. \\
SMALL    &  Arbitrary small number (1.E-10) \\
RADDEG   &  Radian to degree conversion factor \\
DEGRAD   &  Degree to radian conversion factor \\
\end{tabular}

\vspace{3mm}
\begin{verbatim}
+KEEP, SICNTR.
       COMMON /SICNTR/ NPRTAC, NCHAPA, NACCLE, NPLOWT, NPLCAL
      +,               NPARSC, NEFLOW, NGENPA
      +,               LOCPRT, LOCCPA, LOCMPO, LOCLOWT, LOCLCAL
      +,               LOCRSC, LOCEFL, LOCFIN
*
\end{verbatim}
\vspace{-4mm}
\begin{tabular}{ll}
NGENPA   &  Number of all particles in the event history (if IFHIST = 1) \\
         &  or number of only stable particles (if IFHIST = 0)  \\
NPRTAC   &  Number of particles being accepted for simulation \\
NCHAPA   &  Number of charged particles \\
NACCLE   &  Number od particles in the lego plot \\
NPLOWT   &  Number of particles in the low angle tagger \\
NPLCAL   &  Number of particles in the the low angle luminosity calorimeter \\
NPARSC   &  Number of particles being rescued \\
NEFLOW   &  Number of energy flow objects \\
LOCPRT   &  Start location of particles being accepted for simulation \\
LOCCPA   &  Start location of charged particles \\
LOCMPO   &  Start location of particles in the lego plot \\
LOCLOWT  &  Start location of particles in the low angle tagger \\
LOCLCAL  &  Start location of particles in the low angle luminosity calorimeter \\
LOCRSC   &  Start location of particles rescued \\
LOCEFL   &  Start location of energy flow objects \\
LOCFIN   &  Start location of particles for final storage \\
\end{tabular}

\vspace{3mm}
\begin{verbatim}
+KEEP, SIFLAG.
       COMMON /SICONS/ IFPYTH, IFEVTI, IFPLOT, IFGZIO
      +,               IFBEST, IFVECS, IFWDST, IFFORM, IFHIST
      +,               IFAPS,  IFCCD,  IFCOVM, IFIPCT, IFBKGR
      +,               IFLCAL, IFLAT,  IFDEDX
*
\end{verbatim}
\vspace{-4mm}
Flags for enabling resp. disabling of detector components or program
options (see App. C)

\vspace{3mm}
\begin{verbatim}
+KEEP, SIPARA.
       COMMON /SIPARA/ COSCAL, COSECAL, COSHCAL
      +,               EMECAL, EMINHA
      +,               EPIMIS, XMUMIS
      +,               COSCCD, COSAPS, COSVTX
      +,               FIELD,  TPCIRAD, TPCORAD, TPCLEN, QTRKMIS
      +,               TRKEFF, PTMINP, COSTRK, COSTPC
      +,               AEMRES, BEMRES, AHARES, BHARES
      +,               COSLLAC, COSULAC
      +,               EMUDEP, EMUISO
      +,               COSLLAT, COSULAT, EMINLAT
      +,               ALATRES, BLATRES, CLATRES, DLATRES
*
\end{verbatim}
\vspace{-4mm}
Detector parameters (see App. A)

\vspace{3mm}
\begin{verbatim}
+KEEP, PUCPPP.
       PARAMETER (MAXOBJ=500)     ! max. # of e_flow objects (=LOCFIN-LOCEFL)
       COMMON /PUCPPP/PVECP
       DOUBLE PRECISION PVECP(5, 2*MAXOBJ)
       REAL \hspace{29mm} VECP(10, 2*MAXOBJ)
       DIMENSION \hspace{17mm} IVECP(10, 2*MAXOBJ)
       EQUIVALENCE  (VECP(1, 1), PVECP(1, 1))
       EQUIVALENCE (IVECP(1, 1),  VECP(1, 1))
*
\end{verbatim}
\vspace{-4mm}
Working array for VECSUB package

\vspace{3mm}
\begin{verbatim}
+KEEP, PWCPPP.
       PARAMETER (NRAWS = 2000)               ! working array (=LOCFIN)
       PARAMETER (NLOC = 10)
       COMMON /PWCPPP/ PP(NLOC, NRAWS)
       DIMENSION IPP(NLOC, NRAWS)
       EQUIVALENCE (IPP(1,1), PP(1,1))
*
\end{verbatim}
\vspace{-4mm}
Working array for internal particle storage

\vspace{3mm}
\begin{verbatim}
+KEEP, SILATC.
       parameter (maxlat=100)         ! max. # of particles (=LOCLCAL-LOCLOWT)
       parameter (mxplat=20)          ! max. # of particles/cluster
       common /silatc/ numblat(maxlat), linklat(mxplat, maxlat)
      +,               efrclat(mxplat, maxlat)
*
\end{verbatim}
\vspace{-4mm}
Arrays for particle information in the LAT

\vspace{3mm}
\begin{verbatim}
+KEEP, SILEGC.
       parameter (nte=200,npe=400)    ! number of theta/phi cells in em calo
       parameter (nth=100,nph=200)    ! number of theta/phi cells in had calo
       common /silegc/ ecal(nte,npe),hcal(nth,nph),ical(nte,npe)
*
\end{verbatim}
\vspace{-4mm}
Lego arrays for em and hadronic calorimeter

\vspace{3mm}
\begin{verbatim}
+KEEP, SIXREF.
       parameter (mxcel=4000,mxpar=20)
       common /sixref/ ncell,idcel(mxcel),npart(mxcel)
      +,               epart(mxcel,mxpar),idpar(mxcel,mxpar)
      +,               iadre(nte,npe),iadrh(nth,nph)
*
\end{verbatim}
\vspace{-4mm}
Cross reference between calorimeter cell contens and generator particles

\vspace{3mm}
\begin{verbatim}
+KEEP, SICLUC.
       parameter (maxclu=500), maxcel=8000)
       common /sicluc/ nclu(maxclu),eclu(maxclu),tclu(maxclu)
      +,               pclu(maxclu),eecl(maxclu),ehcl(maxclu)
      +,               listc(maxcel,maxclu),ntclu)
*
\end{verbatim}
\vspace{-4mm}
Storage for the output of the cluster algorithm

\vspace{3mm}
\begin{verbatim}
+KEEP, SICLAC.
       common /siclac/ npar(maxclu),listp(mxpar,maxclu)
      +,               ppar(3,maxclu),epar(mxpar,maxclu)
*
\end{verbatim}
\vspace{-4mm}
Cross reference between clusters and generator particles

\vspace{3mm}
\begin{verbatim}
+KEEP, SIFLOC.
       parameter (maxflo=500)       ! max. # of e_flow objects (=LOCFIN-LOCEF)
       common /sifloc/ nflo                       ! number of objects
      +,       ista(maxflo),ityp(maxflo)          ! status and type
      +,       lgen(mxpar,maxflo),ngen(maxflo)    ! mothers
      +,       frac(mxpar,maxflo)                 ! fraction per mother
      +,       pbst( 6,maxflo)                    ! best estimate
      +,       ptrk(15,maxflo)                    ! tracker
      +,       peca( 5,maxflo)                    ! ecal
      +,       phca( 5,maxflo)                    ! hcal
      +,       pmus( 6,maxflo)                    ! muon system
      +,       ltrk(mxpar,maxflo),ntrk(maxflo)    ! track list
      +,       lcal(mxpar,maxflo),ncal(maxflo)    ! cluster list
      +,       lmus(mxpar,maxflo),nmus(maxflo)    ! muon list
*
\end{verbatim}
\vspace{-4mm}
Storage of the results of the energy flow algorythm

\vspace{3mm}
\begin{verbatim}
+KEEP, SIBOUT.
       PARAMETER (NPLUND=1000, NWLUND=13)
       COMMON /SIBOUT/ NGENPA, BOUT(NPLUND,NWLUND)
       DIMENSION IBOUT(NPLUND,NWLUND)
       EQUIVALENCE (BOUT(1,1), IBOUT(1,1))
*
\end{verbatim}
\vspace{-4mm}
Storage array of generator particle information

\vspace{3mm}
\begin{verbatim}
+KEEP, SIENFL.
       PARAMETER (NMXEFL=500)                      ! max. # of e_flows
       PARAMETER (NWENFL=1000)                     ! max. # of words/e_flow
       COMMON /SIENFL/ NENFLO, CM(NMXEFL,NWENFL)
      +,               NGENNW(NMXEFL), NTRKNW(NMXEFL)
      +,               NCALNW(NMXEFL)
      +,               NMUSNW(NMXEFL), NLATNW(NMXEFL)
      +,               NLCANW(NMXEFL)
       DIMENSION ICM(NMXEFL,NWENFL)
       EQUIVALENCE (CM(1,1), ICM(1,1))
*
\end{verbatim}
\vspace{-4mm}
Storage array of energy flow objects (see App.D)

\vspace{3mm}
\begin{verbatim}
+KEEP, SIBSTC.
       PARAMETER (NMXBST=2010)
       COMMON /SIBSTC/ BEST(NMXBST)
       DIMENSION IBEST(NMXBST)
       EQUIVALENCE (BEST(1), IBEST(1))
*
\end{verbatim}
\vspace{-4mm}
Storage array of best energy flow object information (see App. E)

\vspace{3mm}
\begin{verbatim}
+KEEP, SIERRTRK.
       parameter (maxtrk=300)             ! max. # of tracks (= LOCMPO-LOCCPA)
       common /sierrtrk/ rphitrk(maxtrk), rthetrk(maxtrk),rptitrk(maxtrk)
      +,                 dphitrk(maxtrk), dthetrk(maxtrk),dptotrk(maxtrk)
*
\end{verbatim}
\vspace{-4mm}
Track resolution and error for 1/$p_{T}$, $\theta$ and $\phi$

\vspace{3mm}
\begin{verbatim}
+KEEP, SIMXEL.
       parameter (maxcpa=300)             ! max. # of tracks (= LOCMPO-LOCCPA)
       common /simxel/ covxel(15,maxcpa), icovmtx(maxcpa)
*
\end{verbatim}
\vspace{-4mm}
Elements of the covariance matrix

\vspace{3mm}
\begin{verbatim}
+KEEP, IDEDXPAR.
       COMMON /IDEDXPAR/ DEDXPAR(5)
       DIMENSION IDEDXPAR(5)
       EQUIVALENCE (DEDXPAR(1), IDEDXPAR(1))
*
\end{verbatim}
\vspace{-4mm}
Data card parameters for $dE/dx$ estimation

\vspace{3mm}
\begin{verbatim}
+KEEP, SIDEDXC.
       parameter (maxtpc=300)             ! max. # of tracks (= LOCMPO-LOCCPA)
       parameter (maxpnt=1000)            ! max. # of space points per track
       common /sidedxc/ RATIO(maxtpc)
      +,                TRK_P(maxtpc,maxpnt,3), STEP, INDCUT
      +,                RLENGTH(maxtpc), N_POINTS(maxtpc), P_IMP(maxtpc)
      +,                P_TTT(maxtpc), ICHRG(maxtpc)
      +,                RCLOSE(maxtpc), ICLOSE(maxtpc), IDEDX(maxtpc)
      +,                DEDXWGT(5,maxtpc), DEDXNORM(5,maxtpc)
*
\end{verbatim}
\vspace{-4mm}
Parameters for $dE/dx$ estimates

\vspace{3mm}
\begin{verbatim}
+KEEP, DEDXPAR.
       REAL       XI,AKAPPA,XA,AA,PFAC,EXPB,PRES
       PARAMETER (XI     =  0.4720818 ,
      +           AKAPPA = 11.862274  ,
      +           XA     =  2.2334855 ,
      +           AA     =  0.1717262 ,
      +           PFAC   =  1.1292883 ,
      +           EXPB   =  2.3909695 ,
      +           PRES   =  1.0000000 )
       REAL       RESOL,RESEXP,SPLEXP
       PARAMETER (RESOL  =  0.52 ,
      +           RESEXP = -0.47 ,
      +           SPLEXP =  0.32 )
       REAL       RSPLMN,DEMX
       PARAMETER (RSPLMN = 0.20 ,
      +           DEMX   = 250. )
*
\end{verbatim}
\vspace{-4mm}
Bethe-Bloch, resolution and protection parameters for $dE/dx$ estimates

\vspace{3mm}
\begin{verbatim}
+KEEP, DXPROB.
       REAL            DEDXE,DDEDXE,DEDXM,DDEDXM,
      +                PE,PM,DPM,XM,Q,WEIGHT
       INTEGER         METHOD
       COMMON /DXPROB/ DEDXE,DDEDXE,DEDXM,DDEDXM,
      +                PE,PM,DPM,XM,Q,WEIGHT,METHOD
*
\end{verbatim}
\vspace{-4mm}
dE/dx weight and probability values

\vspace{3mm}
\begin{verbatim}
+KEEP, SILINK.
       COMMON /SILINK/ LVECP(NMXEFL)
*
\end{verbatim}
\vspace{-4mm}
Link to line number of particle in PYTHIA record
if the VECSUB package is enabled

\vspace{7mm}
For the CVS structure the common blocks are given in the simdet/include
subdirectories with the replacement of e.g. +KEEP, SIEVNT by
sievnt.inc.

\newpage
\section*{Appendix C: Data cards}

\textbf{This describes the SIMDET V4.01 free format data cards.}

The initialized default values are given in parentheses.
Note that file names should be enclosed in single quotes,
start with a \textbf{point} and end with a \textbf{blank}.
The length of all file names is restriced to \textbf{80 characters}.

Certain obvious logical restrictions apply to the
proper combination of options. For example, option 'PYTH 1'
does not allow for event reading from an external file.

\vspace{7mm}
\begin{tabular}{|l|l|l|l|}
\hline
 Key  & Variable  & Data type  & Meaning  \\
\hline
 NEVT  & NEVENT  & INTEGER  & total number of events to process (D=10)  \\
 PYTH  & IFPYTH  & INTEGER  & 0 = no PYTHIA event generation \\
       &         &          & 1 = event generation by PYTHIA without beamstrahlung \\
       &         &          & 2 = event generation by PYTHIA with beamstrahlung \\
       &         &          &-1 = only \gaga background events (D=1) \\
 EVTI  & IFEVTI  & INTEGER  & 1 = event reading from an external file, \\
       &         &          & written by sr LCWRITE, 0 = no event reading (D=0) \\
 GENF  & LUNGENF & INTEGER  & input file name of generated events, user defined  \\
 WDST  & IFWDST  & INTEGER  & 1 = event writing to an external file, \\
       &         &          & 0 = no event writing (D=1)  \\
 RECF  & LUNRECF & INTEGER  & output file name of processed events, user defined  \\
 FORM  & IFFORM  & INTEGER  & 1 = formatted output structure, \\
       &         &          & 0 = unformatted output structure (D=0)  \\
 GZIO  & IFGZIO  & INTEGER  & 1 = zipped output is enabled, \\
       &         &          & 0 = no zipped otput (D=0)  \\
 IBKG  & IFBKGR  & INTEGER  & 1 = background event reading from an external file, \\
       &         &          & 0 = no background event reading (D=0) \\
 NBKG  & BKGEVT  & REAL     & average number of background events \\
       &         &          & to be overlayed to each physical event (D=0.)\\
 BKGF  & LUNBKGF & INTEGER  & input file name of background events, user defined  \\
 & & & \\
 CLIC  & IFCLIC  & INTEGER  & 0 = CLIC otion disabled \\
       &         &          & 1 = CLIC option enabled without lumi spectrum file \\
       &         &          & 2 = CLIC option enabled with lumi spectrum file (D=0) \\
 CLCF  & LUNCLCF & INTEGER  & input file name of CLIC lumi spectrum file, user defined \\
 & & & \\
 BEST  & IFBEST  & INTEGER  & 1 = best estimates are enabled, \\
       &         &          & 0 = no best estimates (D=0)  \\
 VECS  & IFVECS  & INTEGER  & 1 = filling of the array VECP(I,K) for VECSUB \\
       &         &          & only if IFBEST = 1  \\
       &         &          & 0 = no filling of this array (D=0)  \\
 HIST  & IFHIST  & INTEGER  & 1 = history of the generated event to output file, \\
       &         &          & 0 = only stable generated particles to output file (D=0)  \\
 COVM  & IFCOVM  & INTEGER  & 1 = covariance matrix is enabled,\\
       &         &          & 0 = no covariance matrix (D=0) \\
 IPCT  & IFIPCT  & INTEGER  & 1 = IP beam constraint enabled, \\
       &         &          & 0 = no IP beam constraint (D=0)  \\
\hline
\end{tabular}

\newpage

\begin{tabular}{|l|l|l|l|}
\hline
 Key  & Variable  & Data type  & Meaning  \\
\hline
 PLOT  & IFPLOT  & INTEGER  & 1 = detector response plots are enabled, \\
       &         &          & 0 = no detector response plots (D=0)  \\
 APS   & IFAPS   & INTEGER  & 1 = pixel detector is enabled, \\
       &         &          & 0 = no pixel detector (D=0)  \\
 CCD   & IFCCD   & INTEGER  & 1 = CCD detector is enabled, \\
       &         &          & 0 = no CCD detector (D=1)  \\
 LCAL  & IFLCAL  & INTEGER  & 1 = low-angle calorimeter is enabled, \\
       &         &          & 0 = no low-angle calorimeter (D=1) \\
 LAT   & IFLAT   & INTEGER  & 1 = low-angle tagger (instrumented mask) is enabled, \\
       &         &          & 0 = no low-angle tagger (D=1)  \\
 DEDX  & IFDEDX  & INTEGER  & 1 = dEdx information enabled, \\
       &         &          & 0 = no dEdx information (D=0) \\
       &         &          & 2. paramter (real): step size [cm] (D=3.0) \\
       &         &          & 3. paramter (real): two-particle seperation\\
       &         &          & distance in the $(r-z)$ projection [cm] (D=1.0) \\
       &         &          & 4. paramter (real): two-particle seperation\\
       &         &          & distance in the $(r-\phi)$ projection [cm] (D=0.22) \\
       &         &          & 5. paramter (integer): cut-off parameter \\
       &         &          & 0 = track comparison is stopped after 5 consecutive steps \\
       &         &          & 1 = complete track comparison (D=0)  \\
 EFTR  & TRKEFF  & REAL  & measurement track efficiency (D=0.99)  \\
 PTTR  & PTMINP  & REAL  & min transverse momentum required [GeV] (D=0.10) \\
 QMTR  & QTRKMIS & REAL  & charge misinterpretation probability (D=0.005)  \\
 EMPH  & EMECAL  & REAL  & min energy required for \\
       &         &       & photon detection [GeV](D=0.200)  \\
 EPIM  & EPIMIS  & REAL  & electron misidentification probability (D=0.002) \\
 AEMR  & AEMRES  & REAL  & first energy resolution parameter for ECAL (D=0.145)  \\
 BEMR  & BEMRES  & REAL  & second (constant) energy resolution  \\
       &         &       & parameter for ECAL (D=0.015)  \\
 EMHA  & EMINHA  & REAL  & min energy required for hadron detection \\
       &         &       & in HCAL, [GeV] (D=0.500)  \\
 AHAR  & AHARES  & REAL  & first energy resolution parameter for HCAL (D=0.554)  \\
 BHAR  & BHARES  & REAL  & second (constant) energy resolution \\
       &         &       & parameter for HCAL (D=0.166) \\
 ELAT  & EMINLAT & REAL  & min energy required for \\
       &         &       & particle detection in LAT [GeV] (D=5.0)  \\
 ALAT  & ALATRES & REAL  & first energy resolution parameter for LAT (D=0.10)\\
 BLAT  & BLATRES & REAL  & second (constant) energy resolution  \\
       &         &       & parameter for LAT (D=0.01)  \\
 CLAT  & CLATRES & REAL  & theta angular resolution in LAT [rad] (D=0.04)  \\
 DLAT  & DLATRES & REAL  & phi angular resolution in LAT [rad] (D=0.262) \\
 & & & \\
 EMMU  & EMUDEP  & REAL  & deposited muon energy in calorimeters [GeV] (D=3.8) \\
 EMIS  & EMUISO  & REAL  & min energy for isolated muons [GeV] (D=5.0)  \\
 XMUM  & XMUMIS  & REAL  & muon misidentification probability (D=0.005) \\
\hline
\end{tabular}

\newpage
\section*{Appendix D: Full output structure}

\textbf{This describes the SIMDET V4.01 standard output structure}.

The output file starts with a single file header of 10 words.
Each event starts with the number of generated particles \textbf{NGENPA},
which is followed by a record of 13 words for each particle.
According to the user's choice, the whole event history
(\textbf{HIST 1}) or only the stable particles 
which might give a response in the detector are covered.

The number of energy flow objects, \textbf{NEFLOW}, then follows.
For each object, blocks of records follow, describing status
information, best estimates, generator information, charged particles,
calorimeter clusters and muons from the muon system, in this order.

\subsubsection*{File header record:}
 
\begin{tabular}{|l|l|l|l|}
\hline
 Offset  & Variable name  & Data type  & Meaning  \\
\hline
 1  & INSTATE  & INTEGER  & ISR flag, PYTHIA default:  \\
 & & & 0 = no radiation, \\
 & & & 1 = ISR \& Beamstrahlung, \\
 & & & 2 = ISR only, \\
 & & & 3 = Beamstrahlung  \\
 2  & SQRTS  & REAL  & nominal collider cms energy [GeV]  \\
 3  & SIGMA  & REAL  & Cross section [fb], if provided  \\
 4  & IDCLASS  & INTEGER  & Reaction identifier, user defined  \\
 5  & NEVENT & INTEGER  & Number of events to be processed \\
 6  & IFBEST & INTEGER  & Flag for best estimates  \\
 7  & IFFORM & INTEGER  & Flag for formatted output \\
 8  & IFDEDX & INTEGER  & Flag for dEdx \\
 9  & IFHIST & INTEGER  & Flag for complete (restricted) event history \\
 10 & IFBKGR & INTEGER  & Flag for background events \\
\hline
\end{tabular}

\vspace{3mm}
\noi This record is written once per file.

\subsubsection*{Generated particle record:}
 
\begin{tabular}{|l|l|l|l|}
\hline
 Offset  & Variable name  & Data type  & Meaning  \\
\hline
 1  & STATUS  & INTEGER  & 0 = no detector response, 1 = detector response  \\
 2  & ID  & INTEGER  & particle code according to PYTHIA convention  \\
 3  & line  & INTEGER  & line number of particle in the PYTHIA record  \\
 4  & Px  & REAL  & x-component of momentum [GeV]  \\
 5  & Py  & REAL  & y-component of momentum [GeV]  \\
 6  & Pz  & REAL  & z-component of momentum [GeV]  \\
 7  & E  & REAL  & Energy [GeV]  \\
 8  & m  & REAL  & mass of particle [GeV]  \\
 9  & Q  & REAL  & charge of particle  \\
 10  & x  & REAL  & x-component of vertex [mm]  \\
 11  & y  & REAL  & y-component of vertex [mm]  \\
 12  & z  & REAL  & z-component of vertex [mm]  \\
 13  & Time  & REAL  & time of production [mm/c] \\
\hline
\end{tabular}

\vspace{3mm}
\noi This record is repeated \textbf{NGENPA} times.

\subsubsection*{Energy flow record 1: status of energy flow object}

\begin{tabular}{|l|l|l|l|}
\hline
 Offset  & Variable name  & Data type  & Meaning  \\
\hline
 1  & STATUS  & INTEGER  & $<$ 0: invalid object, \\
 & & & 1=charged object, \\
 & & & 2=neutral object, \\
 & & & 3=composite object \\
 2  & Type  & INTEGER  & particle code according to PYTHIA convention,  \\
 & & & 999 = cluster of unresolvable particles  \\
 3  & NGEN  & INTEGER  & number of generator particles contributing  \\
 & & & to the energy flow object  \\
 4  & NTRK  & INTEGER  & number of charged particles contributing \\
 & & & to the energy flow object  \\
 5  & NCAL  & INTEGER  & number of clusters contributing to the energy flow object  \\
 6  & NMUS  & INTEGER  & number of muons contributing to the energy flow object \\
\hline
\end{tabular}

\vspace{3mm}
\noi This record and the following blocks are repeated \textbf{NEFLOW}
times.

\subsubsection*{Energy flow record 2: best estimate for object's energy and direction}

\begin{tabular}{|l|l|l|l|}
\hline
 Offset  & Variable name  & Data type  & Meaning  \\
\hline
 1  & Px  & REAL  & x-component of momentum [GeV]  \\
 2  & Py  & REAL  & y-component of momentum [GeV]  \\
 3  & Pz  & REAL  & z-component of momentum [GeV]  \\
 4  & E  & REAL  & Energy [GeV]  \\
 5  & m  & REAL  & mass of particle [GeV]  \\
 6  & Q  & REAL  & charge of particle \\
\hline
\end{tabular}

\vspace{3mm}
\noi This record is written once per energy flow object.

\subsubsection*{Energy flow record 3: generator particle contributing to energy flow object}

\begin{tabular}{|l|l|l|l|}
\hline
 Offset  & Variable name  & Data type  & Meaning  \\
\hline
 1  & link  & INTEGER  & line number of particle in the PYTHIA record  \\
 2  & Efrac  & REAL  & energy fraction  \\
\hline
\end{tabular}

\vspace{3mm}
\noi This record is repeated \textbf{NGEN} times.

\subsubsection*{Energy flow record 4: charged particle tracks that are part of the object}

\begin{tabular}{|l|l|l|l|}
\hline
 Offset  & Variable name  & Data type  & Meaning  \\
\hline
 1  & P  & REAL  & absolute value of momentum [GeV]  \\
 2  & Theta  & REAL  & polar angle [radian]  \\
 3  & Phi  & REAL  & azimuth angle [radian]  \\
 4  & Q  & REAL  & charge of particle  \\
 5  & Imp(R,phi)  & REAL  & transverse impact parameter [cm] \\
 6  & Imp(R,z)  & REAL  & longitudinal impact parameter [cm] \\
 7  & cov. matrix  & REAL  & xy-xy element  \\
 8  & cov. matrix  & REAL  & xy-z element  \\
 9  & cov. matrix  & REAL  & z-z element   \\
 10  & cov. matrix  & REAL  & xy-theta element \\
 11  & cov. matrix  & REAL  & z-theta element  \\
 12  & cov. matrix  & REAL  & theta-theta element \\
 13  & cov. matrix  & REAL  & xy-phi element \\
 14  & cov. matrix  & REAL  & z-phi element  \\
 15  & cov. matrix  & REAL  & phi-theta element \\
 16  & cov. matrix  & REAL  & phi-phi element \\
 17  & cov. matrix  & REAL  & xy-1/p element \\
 18  & cov. matrix  & REAL  & z-1/p element \\
 19  & cov. matrix  & REAL  & theta-1/p element \\
 20  & cov. matrix  & REAL  & phi-1/p element \\
 21  & cov. matrix  & REAL  & 1/p-1/p element  \\
 22  & dEdx         & REAL  & signed electron probability \\
 23  & dEdx         & REAL  & signed muon probability     \\
 24  & dEdx         & REAL  & signed pion probability     \\
 25  & dEdx         & REAL  & signed kaon probability     \\
 26  & dEdx         & REAL  & signed proton probability   \\
 27  & dEdx         & REAL  & normalised electron dEdx value \\
 28  & dEdx         & REAL  & normalised muon dEdx value \\
 29  & dEdx         & REAL  & normalised pion dEdx value \\
 30  & dEdx         & REAL  & normalised kaon dEdx value \\
 31  & dEdx         & REAL  & normalised proton dEdx value \\
\hline
\end{tabular}

\vspace{3mm}
\noi This record is repeated \textbf{NTRK} times.

\subsubsection*{Energy flow record 5: calorimeter cluster 
(ECAL/HCAL/LAT/LCAL) that are part of the object}

\begin{tabular}{|l|l|l|l|}
\hline
 Offset  & Variable name  & Data type  & Meaning  \\
\hline
 1  & E  & REAL  & energy [GeV]  \\
 2  & Theta  & REAL  & polar angle [radian]  \\
 3  & Phi  & REAL  & azimuth angle [radian]  \\
 4  & Time  & REAL  & timing information  \\
 5  & C(em)  & REAL  & probability of being consistent \\
 & & & with an electromagnetic particle  \\
 6  & E  & REAL  & energy [GeV]  \\
 7  & Theta  & REAL  & polar angle [radian]  \\
 8  & Phi  & REAL  & azimuth angle [radian]  \\
 9  & Time  & REAL  & timing information  \\
 10  & C(mip)  & REAL  & probability of being consistent \\
 & & & with a minimum ionizing particle \\
\hline
\end{tabular}

\vspace{3mm}
\noi This record is repeated \textbf{NCAL} times.

\subsubsection*{Energy flow record 6: muons from the muon system that are part of the object}

\begin{tabular}{|l|l|l|l|}
\hline
 Offset  & Variable name  & Data type  & Meaning  \\
\hline
 1  & E  & REAL  & energy [GeV]  \\
 2  & Theta  & REAL  & polar angle [radian]  \\
 3  & Phi  & REAL  & azimuth angle [radian]  \\
 4  & Q  & REAL  & charge of particle  \\
 5  & Time  & REAL  & timing information  \\
 6  & C(punch)  & REAL  & probability of being consistent \\
& & & with an punch-through object \\
\hline
\end{tabular}

\vspace{3mm}
\noi This record is repeated \textbf{NMUS} times.

\newpage
\section*{Appendix E: Restricted output structure}

\textbf{This describes the SIMDET V4.01 restricted output structure},
mainly intended to save disc space for reconstructed events.
The output file starts with a single file header of 10 words.
 Each event starts with the number of energy flow objects,
 \textbf{NEFLOW}.
This is followed by a series of records for each energy flow object,
 which describe only the best estimate of the object's energy 
and direction, in analogy to energy flow record 2 
from the full data format, supplementerd by impact parameter information.

\subsubsection*{File header record:}

\begin{tabular}{|l|l|l|l|}
\hline
 Offset  & Variable name  & Data type  & Meaning  \\
\hline
 1  & INSTATE  & INTEGER  & ISR flag, PYTHIA default:  \\
 & & & 0 = no radiation, \\
 & & & 1 = ISR \& Beamstrahlung, \\
 & & & 2 = ISR only, \\
 & & & 3 = Beamstrahlung  \\
 2  & SQRTS  & REAL  & nominal collider cms energy [GeV]  \\
 3  & SIGMA  & REAL  & Cross section [fb], if provided  \\
 4  & IDCLASS  & INTEGER  & Reaction identifier, user defined  \\
 5  & NEVENT & INTEGER  & Number of events to be processed \\
 6  & IFBEST & INTEGER  & Flag for best estimates  \\
 7  & IFFORM & INTEGER  & Flag for formatted output \\
 8  & IFDEDX & INTEGER  & Flag for dEdx information\\
 9  & IFHIST & INTEGER  & Flag for complete (restricted) event history \\
 10 & IFBKGR & INTEGER  & Flag for background events \\
\hline
\end{tabular}

\vspace{3mm}
\noi This record is written once per file.

\subsubsection*{Energy flow record:}

\begin{tabular}{|l|l|l|l|}
\hline
 Offset  & Variable name  & Data type  & Meaning  \\
\hline
 1  & Px  & REAL  & best estimate for x-component of momentum [GeV]  \\
 2  & Py  & REAL  & best estimate for y-component of momentum [GeV]  \\
 3  & Pz  & REAL  & best estimate for z-component of momentum [GeV]  \\
 4  & E  & REAL  & best estimate for energy [GeV]  \\
 5  & m  & REAL  & mass of particle [GeV]  \\
 6  & Q  & REAL  & charge of particle  \\
 7  & Imp(R,phi)  & REAL  & transverse impact parameter, in units of sigma  \\
 8  & Imp(R,z)  & REAL  & longitudinal impact parameter, in units of sigma  \\
 9  & NTRK  & INTEGER  & number of charged particles contributing \\
& & & to the energy flow object  \\
 10  & type  & INTEGER  & particle code according to PYTHIA convention, \\
 & & & 999 = cluster of unresolvable particles \\
\hline
\end{tabular}

\vspace{3mm}
\noi In cases of NTRK $>$ 1, the largest impact parameters are recorded.
 This record is repeated \textbf{NEFLOW} times.


\newpage
\begin{thebibliography}{29}

\bibitem{tdr} F. Richard, J.R. Schneider, D. Trines and A. Wagner
Edts., TESLA: The Superconducting Electron-Positron linear Collider
with an Integrated X-Ray Laser Laboratory.
Technical Design Report, Part IV A Detector for TESLA, DESY 2001-011
and ECFA 2001-209 (2001)

\bibitem{geant} GEANT, Detector Description and Simulation Tool, CERN Program
Library Long Write-up W5013 (1994)

\bibitem{brahms} T. Behnke, G. Blair, K. M\"onig and M. Pohl, BRAHMS -
Version 1.00, a Monte Carlo Program for a Detector at a 800 GeV Linear
Collider, November 6, 1998, 
\verb+http://www.hep.ph.rhbnc.ac.uk/~blair/detsim/brahms.html+   

\bibitem{Ohl} T. Ohl, IKDA  96/13-rev., July 1996 and
hep-ph/9607454-rev.

\bibitem{Schulte} D. Schulte and 
\verb+/afs/cern.ch/eng/tev\_phys/user1/hades/src+

\bibitem{Vogt} H. Vogt, Proceedings of the Linear
Collider Workshop 2000, LCWS 2000, Fermilab, October 24-28, 2000

\bibitem{Battaglia} M. Battaglia and
\verb+/afs/cern.ch/eng/tev\_phys/public/simdet/v2.0+

\bibitem{blobel} V. Blobel, Package for constraint least squares
and error propagation, private communication

\bibitem{cdr} R. Brinkmann, G. Materlik, J. Rossbach and A. Wagner
Edts., Conceptual Design of a 500 GeV $\mathrm{e^+e^-}$ Linear
Collider with Integrated X-ray Laser Facility, DESY 1997-048 and ECFA
1997-182 (1997)

\bibitem{version3} M. Pohl and H.J. Schreiber, DESY 99-030, March 1999

\bibitem{hauschild} M. Gruwe, Studies of $dE/dx$ Capabilities of a TPC
for the Future Linear Collider TESLA, LC-DET-2001-043, 2001; \\
M. Hauschild, Proceedings of the Linear
Collider Workshop 2000, LCWS 2000, Fermilab, October 24-28, 2000

\bibitem{vecsub} G. Wormser, DELPHI note, November 16, 1988;
revised by P. Roudeau, September 11, 1989


\end{thebibliography}
\end{document}